\documentclass[%
preprint, onecolumn,
 amsmath,amssymb,
 aps, 
prb,
]{revtex4-2}

\usepackage{graphicx}
\usepackage{dcolumn}
\usepackage{bm}
\usepackage{subcaption}
\usepackage{xcolor}
\usepackage{placeins}
\usepackage{float} 
\usepackage{ulem} 
\usepackage{lineno}
\usepackage{url}
\DeclareUnicodeCharacter{0308}{\"{}}

\begin{document}

\title{A complimentary impedance spectroscopy biosensing method with graphene}

\author{Munis Khan$^{1*}$, Ivan Mijakovic$^{2,3}$ Santosh Pandit$^2$, August Yurgens$^1$}
\affiliation{%
 $^1$Chalmers University of Technology\\
 Department of Microtechnology and Nanoscience\\
 412 96 Göteborg, Sweden\\
 $^2$Chalmers University of Technology\\
 Department of Life Sciences\\
 412 96 Göteborg, Sweden\\
  $^3$Technical University of Denmark\\
 The Novo Nordisk Foundation Center for Biosustainability\\
 2800 Lyngby, Denmark\\
 $^*$email: munis@chalmers.se
}

\begin{abstract}
We present a method where a bioactive functional layer on an electrically conductive thin film with high sheet resistance can be effectively used for complementary electrochemical impedance spectroscopy biosensing. The functional layer's properties, such as double-layer capacitance and charge-transfer resistance, influence the complex impedance of the thin film in direct contact with the layer. These measurements can be performed using a simple low-frequency setup with a lock-in amplifier.  When graphene is used as the resistive thin film, the signal may also include contributions from graphene's quantum capacitance, which is sensitive to charge transfer to and from the graphene. Unlike in traditional graphene biosensors, changes in electrolyte properties over time, such as those caused by the dissolution of ambient gases, do not significantly affect AC measurements. This technique supports biosensor miniaturization, ensures stable operation, and provides reliable biomarker detection with a high signal-to-noise ratio.
\begin{description}
\item[Keywords]
Graphene, Electrochemical Impedance Spectroscopy, Biosensors
 
\end{description}
\end{abstract}

\maketitle


\section{\label{sec:level}Introduction}

Since its discovery, graphene has emerged as one of the most promising two-dimensional (2D) materials for electronics and optoelectronics. This zero-bandgap semiconductor, composed of carbon atoms in a hexagonal lattice \cite{novoselov2004electric}, exhibits excellent electrical properties, including high carrier mobility \cite{tan2007measurement}. These characteristics, combined with its sensitivity to charged species near its vicinity, make graphene an ideal nanomaterial for sensing applications. In particular, graphene field effect transistors (GFET's) have gained significant attention as label-free affinity biosensors due to their ability to translate molecular interactions into electrical signals \cite{hwang2007carrier, ohno2010label}.

Most GFET biosensors are based on liquid-gate transistor configurations, where an electrochemical gate modulates charge transport in an aqueous environment. In this setup, the gate voltage $V_{\mathrm g}$ is applied between the graphene channel and a gate electrode immersed in solution, while a constant current $I_{\mathrm ds}$ flows through the graphene channel. The resulting bell-shaped transfer curve reflects the transition between hole and electron conduction, with the charge-neutrality point (CNP) marking equal carrier populations. The biosensing mechanism is usually attributed to shifts in the CNP, induced by molecular adsorption or molecular binding, resulting in a charge transfer to- or from- graphene. However, detection precision is often hindered by hysteresis due to charge trapping and time drifts \cite{wang2010hysteresis, yu2020effect, wei2020understanding}. Additionally, analyte adsorption introduces charge carrier scattering and increases disorder in the system \cite{chen2014electronic, farid2015detection}.

Electrochemical impedance spectroscopy (EIS) is a valuable method in biosensing, providing detailed information about biorecognition events at the electrode surface. It is used to detect DNA, antigens, antibodies, and bacterial cells \cite{lazanas2023electrochemical, randviir2022review}. EIS probes the signal response over a wide frequency range (~1 mHz - 1 MHz) and is a non-invasive technique that does not disturb the system's steady state. It allows continuous monitoring of biological processes in real-time and detects biological interactions without labels, preserving the natural state of biomolecules. EIS is highly sensitive to changes on the electrode surface, making it ideal for monitoring surface modifications and can detect very low concentrations of analytes, especially with amplification techniques \cite{EIS_ITO_2015}. However, the results are sensitive to experimental setup and conditions, needing careful control and calibration. Additionally, EIS requires specialized and often expensive equipment.

Here, we show that a bio-active functional layer (BFL) on top of an electrically conducting thin film with sufficiently high sheet resistance can conveniently be used as a complementary EIS biosensing method. The functional-layer properties involving the double-layer capacitance and/or charge-transfer resistance is reflected in the complex impedance of the thin-film in direct contact with the layer. A simple low-frequency setup using a lock-in amplifier is sufficient for such measurements.  In the particular case of having graphene as a resistive thin film, the signal can have an additional contribution from the graphene quantum capacitance, which is also sensitive to charge transfer to- and from graphene. Unlike majority of graphene biosensors, the electrolyte properties, which might change with time due to e.g., dissolution of ambient gases in the liquid, are not important in these measurements. Overall, this approach allows for biosensor miniaturization, its drift-free operation, and a reliable detection of biomarkers with high signal-to-noise ratio.

\section{\label{sec:exp}Experimental Methods}
Here, we present the design and fabrication of an antibody-functionalized GFET biosensor that integrates the high specificity of a bio-receptor (human KLK3/prostate-specific antigen (PSA)) with the highly sensitive detection capabilities of a GFET. This biosensor enables rapid, selective, and label-free detection of the PSA antigen. The GFET biosensor, a three-terminal device, utilizes functionalized graphene as a conductive channel between the source and drain electrodes. Fabrication involves multiple stages of conventional microfabrication processing, followed by surface non-covalent functionalization with a molecular linker, which binds to the graphene surface via the $\pi$-$\pi$ stacking interactions \cite{georgakilas2016noncovalent} (see Fig.~\ref{fig_1}b.). The PSA antibody is immobilized onto the graphene channel using carbodiimide cross-linking chemistry \cite{lepvrier2014optimized}. To minimize a non-specific adsorption, the surface of graphene is further blocked with amino-PEG5 alcohol and ethanolamine hydrochloride \cite{gao2016specific, liu2013parts}. 

\subsection{\label{sec:GFET}{GFET fabrication:}} 
\begin{figure}
\includegraphics[width = 16cm]{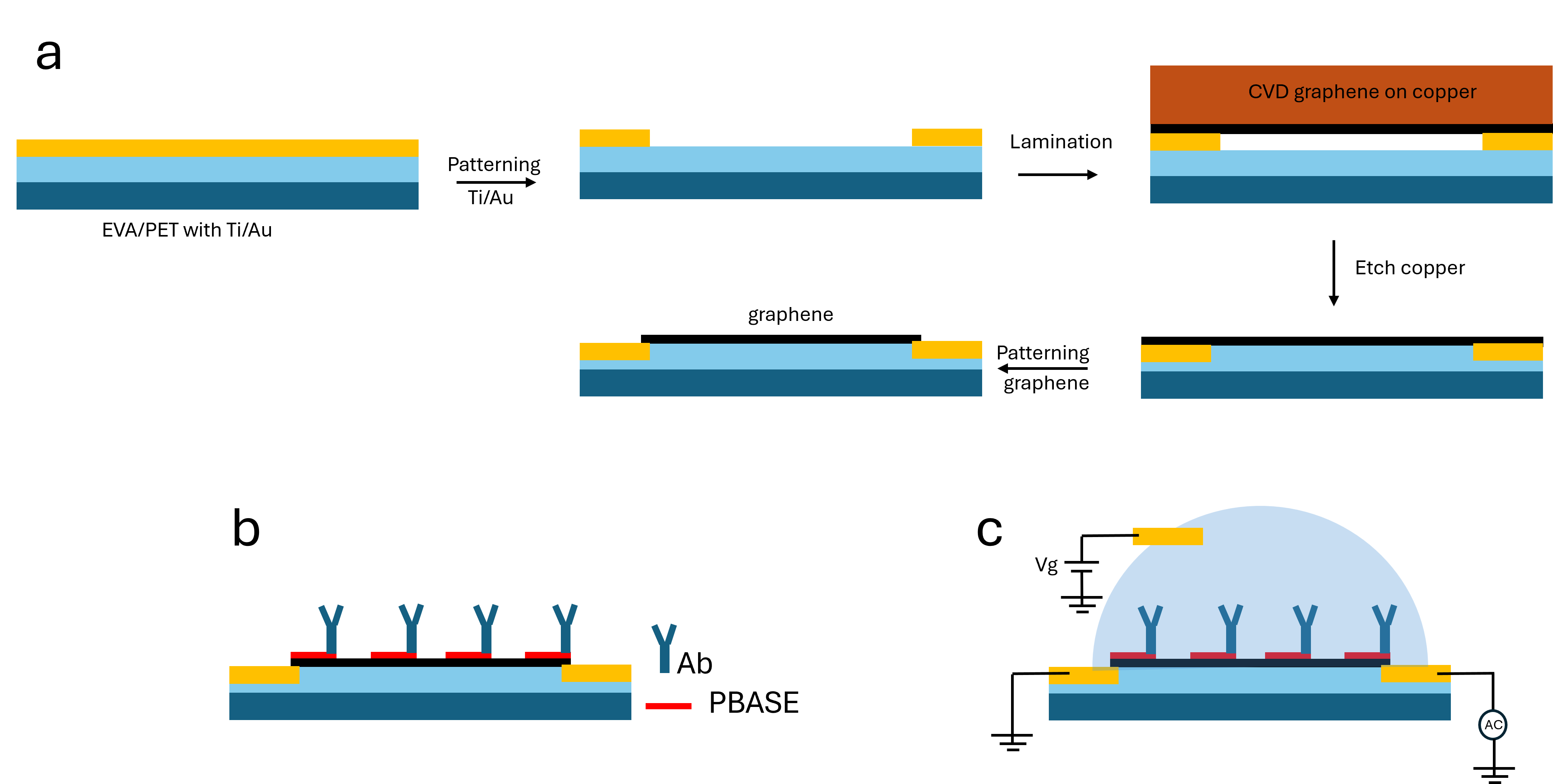}
\caption{\label{fig_1} a) Process flow for graphene transfer from Cu foil to EVA/PET substrate, b) functionalization and c) electrical measurement set-up.}
\end{figure}

The graphene used in the fabrication of GFET is a monolayer chemical-vapor deposited (CVD) graphene on copper foil. We transfer CVD graphene onto ethylene vinyl acetate (EVA)/polyethylene terephthalate (PET) foil by hot-press lamination with prepatterned electrodes on them (the current-bias-, voltage-, and gate electrodes). Copper foil is etched away chemically by diluted nitric acid (10\% $\mathrm{HNO_3}$ in water) follower by graphene patterning using photolithography and oxygen plasma (see Fig.~\ref{fig_1}a).  Finally, a short plastic tube was attached to the GFET chip by using an epoxy glue to form a small well around graphene channel.

\subsection{\label{sec:func}{Functionalization:}} 
Functionalization includes deposition of a single layer of linker molecules of 1-pyrene\\butanoic acid succinimidyl ester (PBASE), and a subsequent deposition of biological reagents for target-molecule capture. First, the GFET devices were incubated with PBASE (5~mM in dimethylformamide (DMF) (Sigma-Aldrich)) for 2~h at room temperature and then rinsed in DMF to remove excess PBASE from the surface before being blow-dried with N$_2$. The linker was then conjugated with Human KLK3/PSA antibody (Sigma-Aldrich). 30-$\mu$L droplets of PSA antibody solution (1~mg/mL) were added to the surface of the chip and left overnight in humid environment at 4~C. The chips were then rinsed in deionized (DI) water and blow-dried with N$_2$. After that, 3~mM PEG5-alcohol (Broadpharm, P-22355) and 3~M ethanolamine hydrochloride (ETA) (Sigma-Aldrich) were used to block the unreacted PBASE molecules. 30-$\mu$L droplets of PEG5-alcohol were added to the surface of the chip and left atop for 1~h at room temperature and then rinsed with DI water and blow-dried with N$_2$. The same procedure was repeated for ETA. Fig.~\ref{fig_1}b shows the functionalized GFET, showing PBASE and antibody attached to the surface of graphene channel.

\subsection{\label{sec:meas}{Electrical measurements:}} 
Following the incubation and cleaning of GFET sensor chip, the four-probe electrical measurements were performed in a low-ionic strength 0.001×PBS solution to avoid the charge screening effect, which reduces the observed signal \cite{palazzo2014detection}. As illustrated in Fig.~\ref{fig_1}c, an AC bias voltage, $V_{\mathrm b}=250$~mV, with the frequency $f=137$~Hz, was applied between the current-bias terminals of the GFET through 1~M$\Omega$ resistor, to get an almost constant AC current $I_{\mathrm b}=250$~nA. The DC liquid-gate voltage $V_{\mathrm g}$ was supplied using Keithley-2604B voltage source meter relative to one of the current-bias electrodes. During the lock-in measurements (SR830, Stanford Research Systems), both the in-phase ($X$) and the quadrature ($Y$) components of the preamplified ($\times 100$) voltage $V$ were recorded. For transfer curves ($R(V_{\mathrm g})$-curve), the GFET's well was filled with $200 \ \mu$L 0.001×PBS solution and the chip was allowed to stabilize for five minutes before the gate voltage was swept at a rate of 10~mV/s and both X and Y components of $V$ were recorded  before- and after introduction of analyte. The transfer curves were cycled between 0 and -0.3V three times to show hysteresis-free behavior. For the time-series measurement, the graphene channel was brought to the peak transconductance with the corresponding $V_{\mathrm g}$. Both X and Y components were recorded vs. time in real time before- and after analyte ($10\ \mu$L of varying concentration) was drop-cast onto the channel. After the tests, the chip was disconnected from the source meter, thoroughly rinsed, refilled with 0.001×PBS, and reconnected to the source meter with the same $V_{\mathrm g}$ and $I_{\mathrm b}$ as before. Once the reconnected chip stabilized, analyte of interest was introduced at different concentrations.  All analytes were prepared in the same 0.001×PBS buffer, to avoid, no matter how small, changes of $pH$-value of the solution in the well upon adding analyte. DC measurements were done by using the Keithley-2604B voltage source or just a battery and measuring the preamplified DC voltage $V_\mathrm{ds}$ by the HP-34401 multimeter.

\section{\label{sec:model}Model} 

Graphene is often mentioned to be very promising material for biosensing applications. And indeed, DC resistance of graphene is very sensitive to charge transfer to- or from graphene channel. This can be used to detect a charge redistribution due to analyte molecular binding in the bio-functional layer in contact with graphene. However, using  AC measurements of the channel resistance, even changes in effective capacitance of the nearby layers can also be detected. Then, instead of graphene, it can be any other conducting and sufficiently thin film from any suitable material with sufficiently high resistivity, e.g., NiCr or TaN. It has long been known that changes in the capacitance of the functional layer can reveal very small concentrations of analyte in EIS method \cite{EIS_ITO_2015}.   

\begin{figure}
	\centering
	\includegraphics[width=1\linewidth]{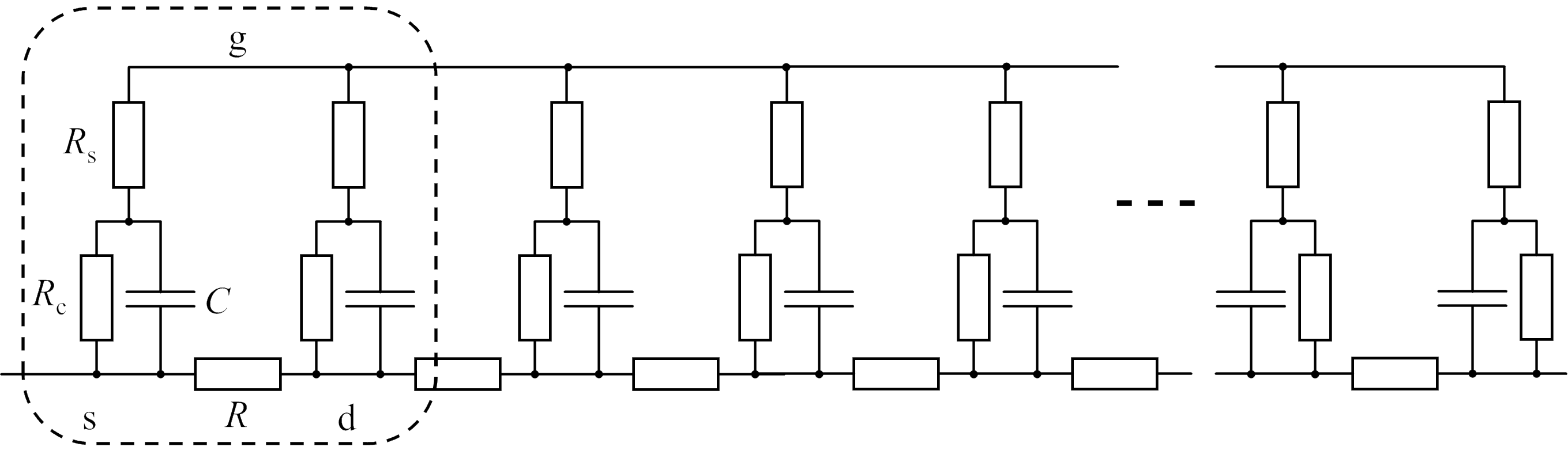}
	\caption{\label{fig_2} Simple model of a liquid-gate biosensor with resistive thin film including the Randles equivalent circuit of electrochemical cell. $C$ is the effective capacitance of double layer plus the layer of functional molecules (for graphene also including its quantum capacitance). $R_c$ is the charge transfer resistance of the double layer, $R_s$ is the resistance of electrolyte, and  $R$ is the thin-film- or graphene resistance.}  
\end{figure}

The GFET with a liquid gate can be modelled by a network of resistive and capacitive components, as shown in Fig.~\ref{fig_2}, representing the distributed Randles equivalent circuit.  For simplicity, we analyze just one element of such a circuit. Here $R_s$ is the solution resistance, $R_c$ is the charge transfer resistance of the double layer, $R$ is the in-plane resistance of graphene, and $\omega$ is the angular frequency. Graphene has a wide electrochemical window in electrolytic environment resulting in a very high $R_c\gg R, R_s$ \cite{zhou2009electrochemical} and can hence be neglected. Since we used a highly diluted buffer (0.001× PBS), $R_s$ is also large, $R_s\ge R$.   

The graphene–electrolyte interface can be modeled as a series combination of three capacitors. The first component is the typical electrical double-layer capacitance $C_{dl}$, which accounts for two layers of ions that are created at the surface of a polarized electrode. Next is the capacitance of the bio-functional layer $C_{f}$, consisting of bulky antibodies and surface-blocking molecules.  In addition, the quantum capacitance $C_{q}$ that accounts for the variation of the density of states with Fermi level must also be considered to fully model the graphene–electrolyte interface. The graphene quantum capacitance is given by Eq.~\ref{eq:Cq} \cite{xia2009measurement}.

\begin{equation}   
	{C}_q= \frac {2e^2}{{\hslash }{v}_F\sqrt {\pi } }\left ( \left | {n}_{g} \right |{+}| {n}_o | \right )^{1/2}
	\label{eq:Cq}
\end{equation}

Typically, $C_\mathrm{dl}\sim C_\mathrm{q}\gg C_\mathrm{f}$ and therefore $C\sim C_\mathrm{f}$. The $X$ and $Y$ components of the voltage drop across resistance $R$ of this simple circuit element, assuming   $R_c=\infty$  and  $R=\epsilon R_s$, $\epsilon \ll 1$ are given by the following equations:  

\begin{equation}
	X =\frac{\epsilon R_s  \left[1+ C^2 R_s^2 \omega^2 (1-\epsilon /2) \right]}{ \left(1+C^2 R_{s}^2 \omega^2 \right)}
	\label{eq:X}
\end{equation}

\begin{equation}
	Y =- \frac{C R_{s}^2 \epsilon^2 \omega}{2 (1+C^2 R_{s}^2 \omega^2 )}
	\label{eq:Y}
\end{equation}

From these simple equations we have discovered an intriguing metric that, by measuring the $X$ and $Y$ components of the AC voltage across a resistive thin film, enables direct access to the capacitance of an adjacent layer on top of the film. Specifically, this metric is represented by the ratio $Y/X^2$ and for small $\omega$, so that for $C^2 R_{s}^2 \omega^2<1$

\begin{equation}
	\frac{Y}{X^2} =- \frac{C \omega}{2 (1+C^2 R_{s}^2 \omega^2)} \rightarrow -\frac{C\omega}{2}.
	\label{eq:YX2}
\end{equation}

The convenience of this metric lies in its low-frequency operation, which allows for the use of inexpensive electronics in practical biosensor devices. In the case of graphene, this metric is also not affected by a baseline time ($t$) drifts in $R(t)$, common in DC measurements of graphene resistance. These drifts are likely caused by changes in graphene doping due to external factors, such as e.g., variations in $pH$ value from the dissolution of $CO_2$ in water. Even $pH$ of buffer solutions is not constant and depends on ionic strength and temperature \cite{enwiki:buffer}. 

\section{\label{sec:results}Results and Discussion}
Real-time biosensing using GFET's typically involves monitoring conductance or resistance over time at a fixed $V_{\mathrm g}$. The optimal response occurs in the high transconductance regions \cite{saltzgaber2013scalable}, but this also amplifies a low-frequency noise, which degrades sensitivity \cite{fu2017biosensing, mavredakis2018understanding}. Furthermore, the baseline drift remains a major challenge in GFET's even in the absence of target molecules \cite{ushiba2020state, miyakawa2021drift, ushiba2024drift}. Conventional methods assume a constant interfacial  capacitance and carrier mobility, but bio-molecules interacting directly with graphene can induce additional charge-carrier scattering, suppressing their mobility \cite{lin2013label}. Practical sensor designs must also account for interfacial capacitance changes upon bio-molecular adsorption \cite{chen2012graphene}.

Here we demonstrate that using AC bias enhances biosensor sensitivity and accuracy by simultaneously capturing both resistance and capacitance changes during bio-molecular interactions. Unlike traditional methods that focus largely on DC conductance, we use AC bias and a low-frequency lock-in amplifier to extract information about dynamic charge transport and electrostatic interactions. Our method provides direct access to three capacitances in series, $C_\mathrm{q}$ \cite{wang2016analytical, wang2016high, pourasl2019quantum},  $C_\mathrm{dl}$  \cite{chen2012graphene, munje2015flexible}, and $C_\mathrm{f}$.

Because of relatively low charge-carrier density in graphene, the quantum capacitance can become important in biosensing \cite{ebrish2014effect, deen2013graphene, ebrish2012operation, hassan2023tuning}. Analyte adsorption alters graphene’s carrier density, leading to measurable shifts in Fermi energy and quantum capacitance \cite{wang2016high, pourasl2019quantum, zhang2017capacitive, xia2009measurement, chen2010graphene, mackin2014current}. Previous studies relied on buried gate electrodes under high-k dielectrics for quantum capacitance measurements \cite{wang2016high, zhang2017capacitive}. These architectures, however, complicate fabrication and limit scalability. Liquid gate eliminates the need for high-k dielectrics, enabling  measurements of $C^{-1}=C_\mathrm{q}^{-1}+C_\mathrm{dl}^{-1}+C_\mathrm{f}^{-1}$ in simple devices while maintaining high sensitivity.

The functional molecules on top of graphene create an ion-permeable charged layer involving accumulation of counter-ions required to maintain the charge neutrality. The concentration difference between the bulk solution and the immobilized ion-permeable layer establishes the Donnan potential \cite{ohshima1985donnan, bergveld1991critical}, which further modifies the electric field between graphene and gate electrode. This additional potential alters the graphene-channel resistance ($R_\mathrm{ch}$), thereby extending sensing beyond the Debye screening length \cite{palazzo2014detection, gao2015general}.

\begin{equation} 
	R =  {R_0 +  R_\mathrm{ch} = R_0 + \frac 1{{ne\mu}} \frac{L}{W}}
	\label{eq:R}
\end{equation} 

The total resistance of the graphene device $R$ can be approximated by Eq.~\ref{eq:R} assuming the constant charge-carrier mobility $\mu$ \cite{kim2009realization}. Here  $n$ is the total charge carrier concentration in graphene, $L$ and $W$ is the length and width of the channel $L/W = 2$ in this work. $R_0$ is the contact resistance and is expected to be zero in the four-probe resistance measurements. Its non-zero value is yet often needed to improve fitting of experimental data by Eq.~\ref{eq:R} and can be taken as a parameter reflecting the small gate dependence of $\mu$ \cite{Jeppson2021}. 
The total charge-carrier density 
$\left( n=\sqrt {n_0^2 +n_g^2} \right)$ is a function of the gate-induced charge density $\left( n_\mathrm{g} =  V_\mathrm{g}C/e \right)$ and the intrinsic carrier concentration ($n_0$), which accounts for charge traps and impurities \cite{adam2007self}. The efficiency of modulation $R$ by the gate voltage is given by the so-called transconductance ($g_m$), which is defined as the derivative of the transfer curve $\Sigma(V_\mathrm{g})$, $\Sigma = 1/R$.

\begin{figure}
\includegraphics[width = 16cm]{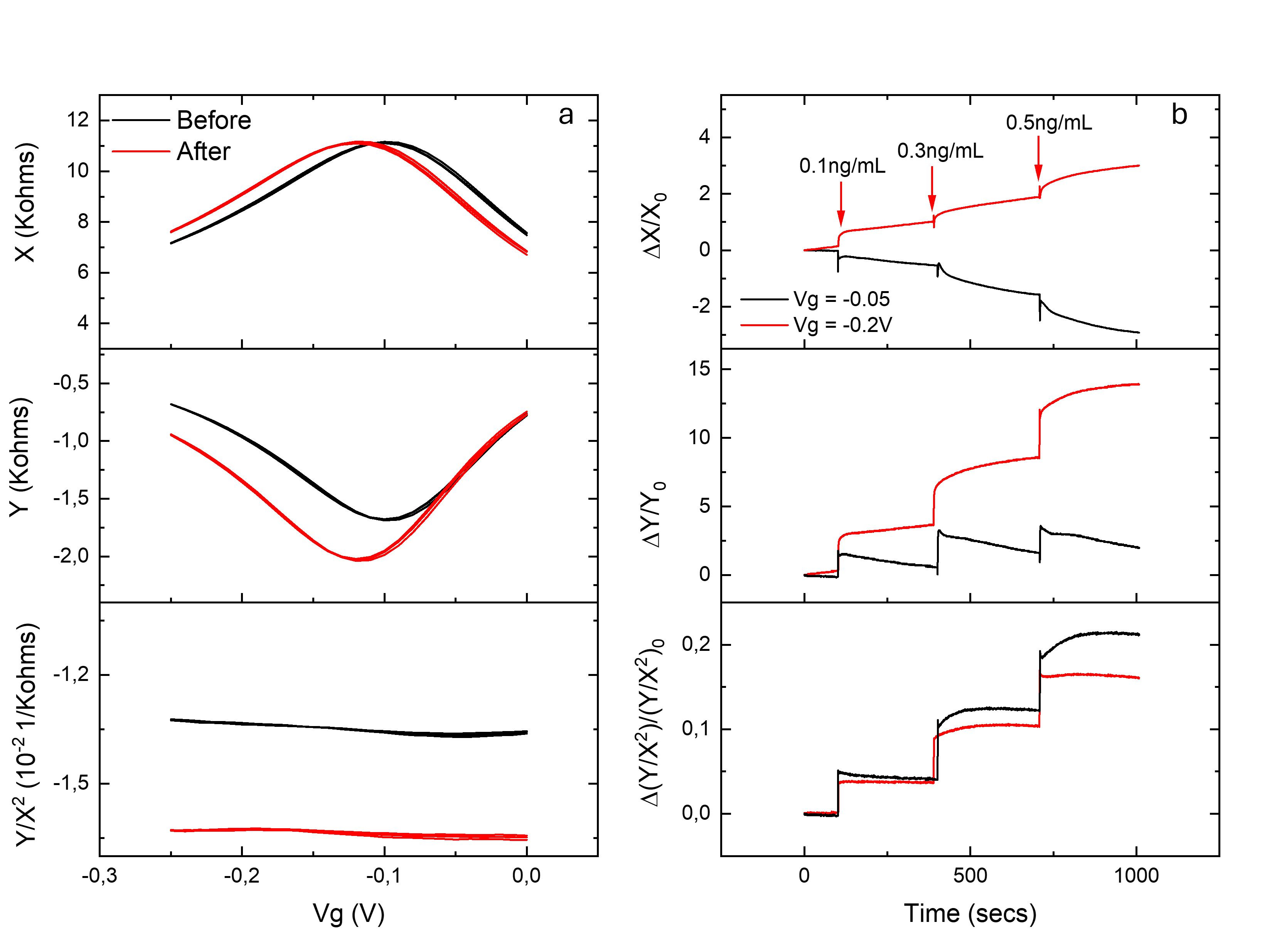}
\caption{\label{fig_3} a) Transfer curves and corresponding b) time series measurements at $V_{\mathrm g}$ -0,05 V and -0,2 V of an PSA antibody functionalized GFET. The red arrows represent the intervals at which 10 $\ \mu$L analyte of varying concentration prepared in 0.001×PBS are dropped on GFET sensors containing 200 $\ \mu$L 0.001×PBS}
\end{figure}

In our lock-in measurements the $X$ component primarily reflected changes in the graphene channel resistance, modulated by the gate voltage ($V_{\mathrm g}$), which alters carrier concentration. Meanwhile, the $Y$ component captured variations in capacitance between the liquid gate and graphene channel, influenced by quantum capacitance and double layer capacitances. 

The measurements of $X$ and $Y$ versus $V_{\mathrm g}$ were performed before and after analyte introduction, as shown in Fig.~\ref{fig_3}a. A shift in both components indicates that the analyte alters the GFET’s electrical properties. A shift in $X$ suggests changes in $R_\mathrm{ch}$, likely due to charge transfer \cite{beraud2021graphene} initiated by analyte binding. This binding modifies the carrier density and shifts the Dirac point, leading to either an increase or decrease in resistance at a constant $V_\mathrm{g}$. A shift in $Y$ includes changes of effective capacitance $C$.  It has been assumed that it is driven by variations in quantum capacitance \cite{deen2013graphene, zhang2017capacitive, hassan2023tuning} and/or modifications in the electric double layer \cite{munje2015flexible, chen2012graphene, randviir2022review}. These changes occur through multiple mechanisms, including ion redistribution in the EDL, which alters its capacitance $C_\mathrm{dl}$, charge transfer, which modifies $C_\mathrm{q}$ \cite{deen2013graphene, zhang2017capacitive, hassan2023tuning}, and changes in the local dielectric constant ($k$), as the analyte displaces water molecules or introduces molecular dipoles, affecting $C_\mathrm{f}$ \cite{thriveni2022advancement}. However, $C_\mathrm{f}$ seems to be the smallest of the three capacitances involved in the measurements and therefore is likely the main sensing readout element in our devices.

As shown in Fig.~\ref{fig_3}a, the shifts in $Y$ were more pronounced than in $X$, suggesting a significant capacitive contribution to the sensing response. We introduced a new metric to further analyze this and plotted $\gamma \equiv Y/X^2$ versus $V_\mathrm {g}$ before- and after analyte introduction. This ratio enhances sensitivity to capacitive effects, effectively isolating them from resistive contributions (see Eg.~\ref{eq:YX2}). Indeed, $Y/X^2$ is nearly constant over the entire $V_\mathrm{g}$ range, while $X(V_\mathrm{g})$ and $Y(V_\mathrm{g})$ are strongly non-linear functions (see Fig.~\ref{fig_3}a). This indicates that $Y/X^2$ is convenient for singling out capacitive changes affected by subtle bio-molecular binding effects, which makes it highly effective for biosensing applications. This parameter provides a stable and reliable detection metric, allowing for better analyte differentiation compared to conventional methods.

To validate this metric further, we performed time ($t$) series measurements (Fig.~\ref{fig_3}b) by fixing $V_\mathrm{g}$ at the peak transconductance points and continuously monitoring $X$, $Y$, and $Y/X^2$ as the analyte was introduced. The GFET was exposed to different analyte concentrations at some time intervals, indicated by the red vertical arrows in Fig.~\ref{fig_3}b. The electrical response was normalized to the values at $t=0$, i.e., $\Delta X/X_0 = (X-X_0)/X_0$, where $X_0=X(t=0)$, etc. A significant drift in $X$ is seen, which obscures the response to adding the analyte.  On the contrary, $Y/X^2$ provided a stronger response, a higher signal-to-noise ratio (SNR), and minimal dependency on $V_\mathrm{g}$, ensuring consistent results across different operating points.

\begin{figure}
\includegraphics[width = 16cm]{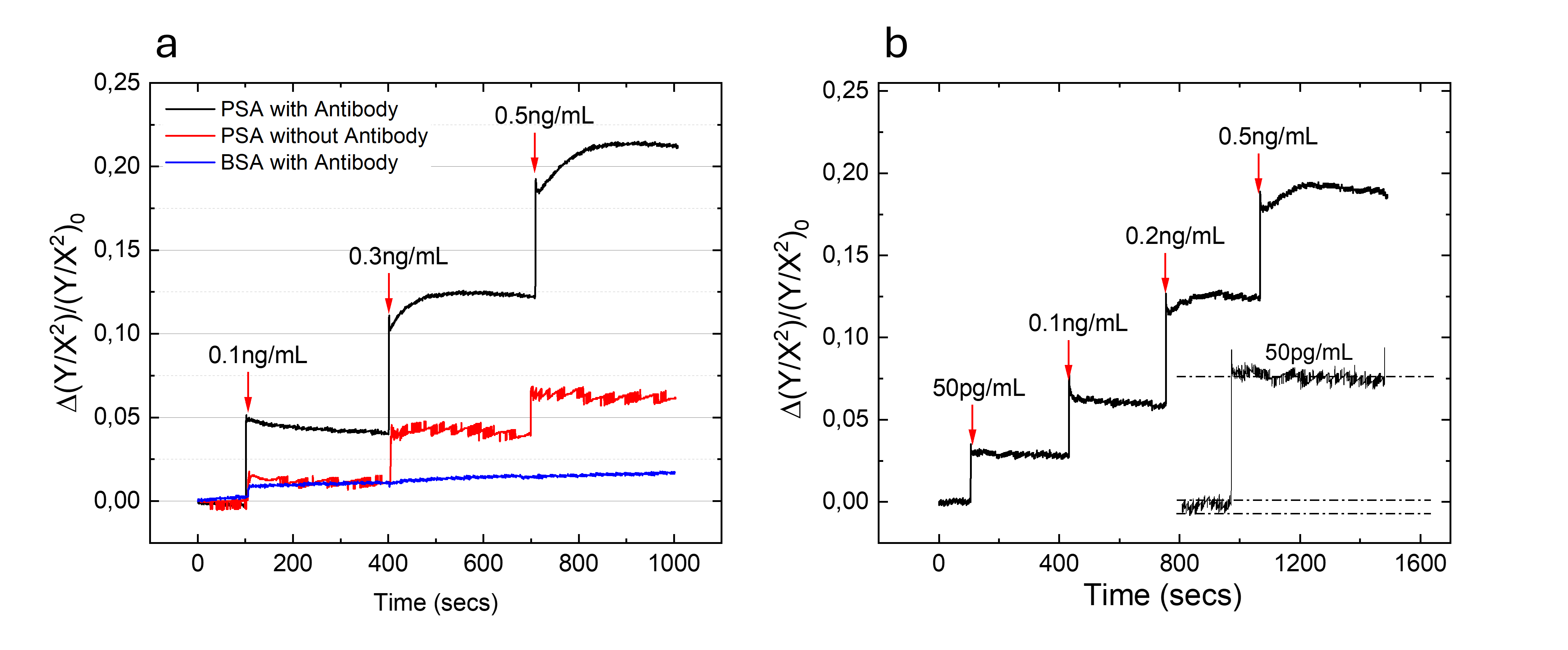}
\caption{\label{fig_4}a) Selective response of a PSA antibody functionalized GFET towards target analyte PSA (black line) and non-target analyte BSA (blue line), also shown is a response of a non-functionalized (without PSA antibody) to target analyte PSA (red line). b) Real time response of a PSA antibody functionalized GFET towards varying concentrations of target analyte PSA, inset shows a SNR to the  50~pg/mL concentration. The red arrows represent the intervals at which 10~$\mu$L analyte of varying concentration prepared in 0.001×PBS are dropped in GFET's well filled with 200~$\mu$L 0.001×PBS}
\end{figure}

\begin{figure}
\includegraphics[width = 16cm]{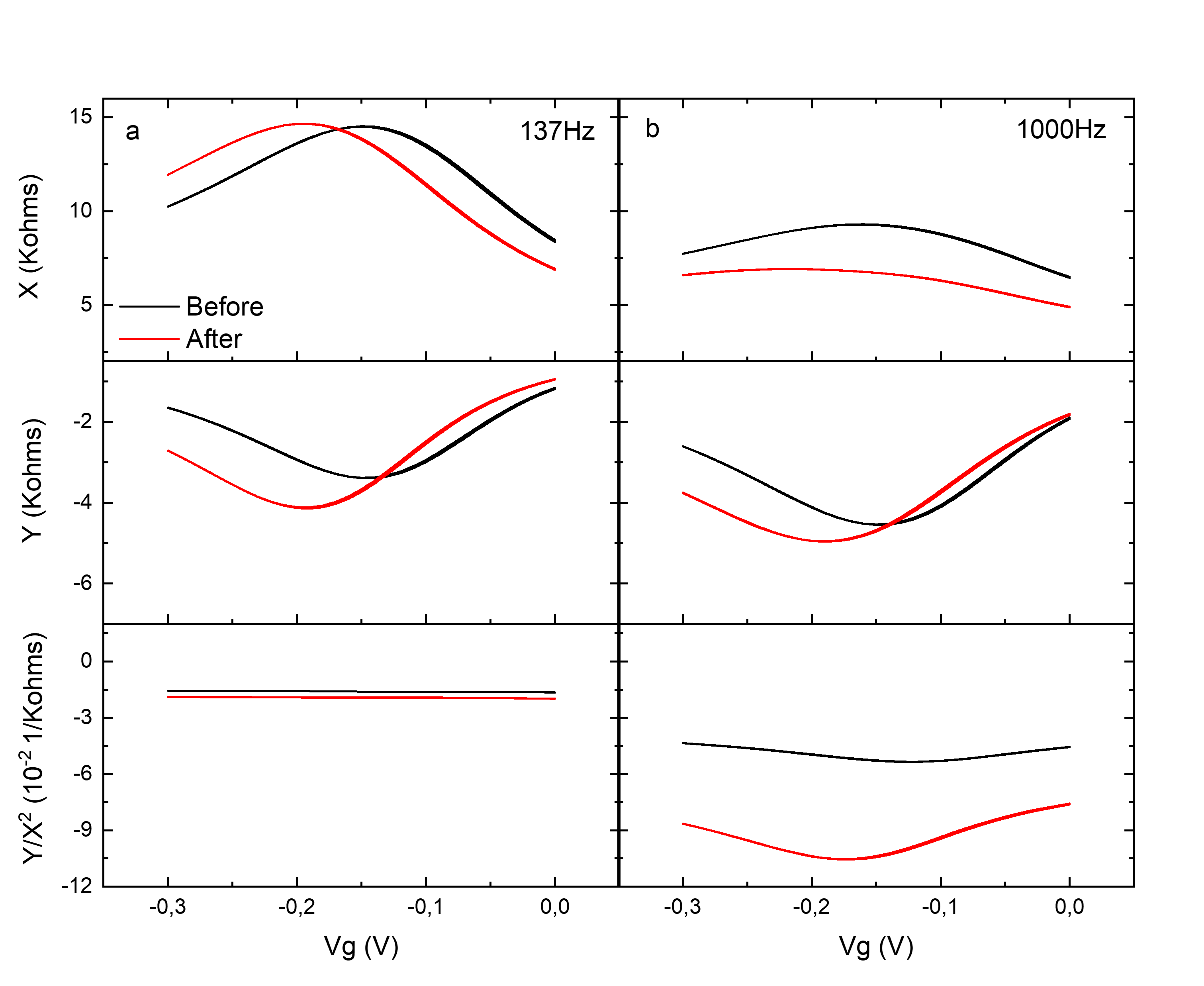}
\caption{\label{fig_5}  $X$, $Y$, and $Y/X^2$ vs $V_\mathrm{g}$ at a) 137~Hz and 1000~Hz before and after target analyte.}
\end{figure}

\begin{figure}
\includegraphics[width = 16cm]{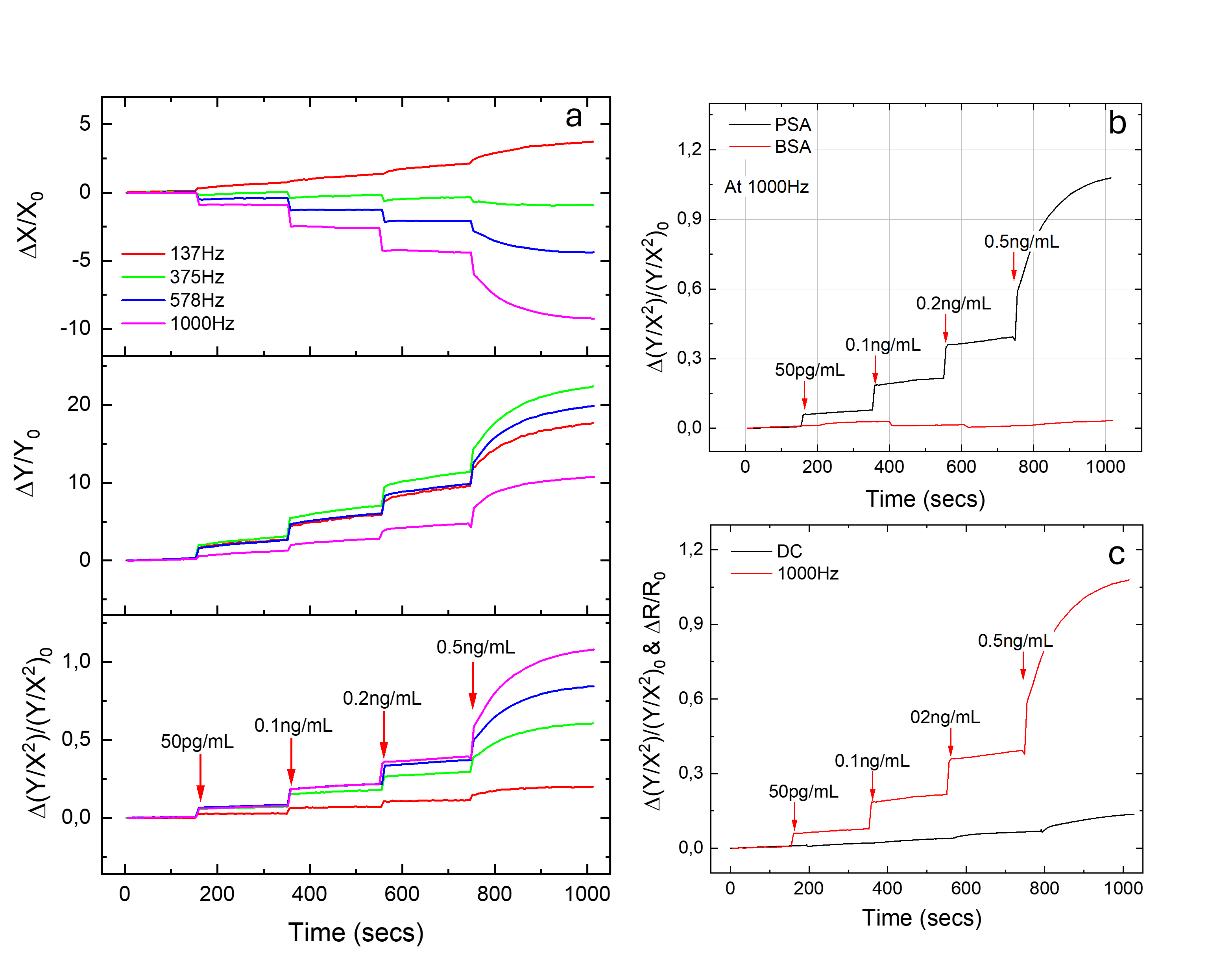}
\caption{\label{fig_6}a) Real time response of a PSA antibody functionalized GFET towards varying concentrations of target analyte PSA at different frequencies. b) Normalized $Y/X^2$ response of target (PSA) and non-target (BSA) at 1000~Hz drive frequency. c) Normalized DC resistance ($\Delta R/R_0$) vs. normalized $Y/X^2$ response of target (PSA) analyte. The red arrows represent the intervals at which 10~$\mu$L analyte of varying concentration prepared in 0.001×PBS are dropped into GFET-sensor well initially filled with 200~$\mu$L 0.001×PBS}
\end{figure}

We also conducted control experiments to check how specific is the response of the functionalized GFET to the target analyte by making tests with a non-target  protein, bovine serum albumin (BSA) (Fig.~\ref{fig_4}a). The red arrows in Fig.~\ref{fig_4}a indicate the time intervals when a specific concentration of either the target protein (PSA) or the non-target protein (BSA) was introduced onto the GFET (refer to the Experimental Methods section). As evident in Fig.~\ref{fig_4}a, the GFET functionalized for PSA exhibits a distinct response only to the target analyte (PSA), confirming its specificity.

To further validate our results, we also tested a non-functionalized GFET (without PSA antibodies) by exposing it to varying concentrations of PSA antigen. The results were then compared with those from the functionalized GFET (with PSA antibodies), as shown in Fig.~\ref{fig_4}a. While a small shift in response was observed for the non-functionalized GFET, the response of the functionalized GFET to PSA at the same concentrations was significantly larger. The minor response in the non-functionalized GFET is likely due to non-specific bindings of PSA to the graphene surface. 

To evaluate the sensitivity of the GFET biosensor, we conducted real-time measurements by introducing lower concentrations of the target analyte (Fig.~\ref{fig_4}b). The experiment began with the addition of 10~$\mu$L of a 50~pg/mL PSA solution into 200~$\mu$L of buffer into GFET's well. This concentration corresponds to 1.6 pM, assuming the molecular weight of PSA antigen as $30\ 000$~Da. A significant response was observed with a high signal-to-noise ratio (see the inset of Fig.~\ref{fig_4}b), demonstrating the sensor's ability to detect PSA at extremely low concentrations. For prostate cancer, a PSA level greater than 4.0 ng/mL is typically regarded as abnormal, potentially leading to a recommendation for a prostate biopsy. However, since PSA levels naturally rise with age, some doctors use a higher threshold (e.g., 5 ng/mL) for older men and a lower threshold (e.g., 2.5 ng/mL) for younger men \cite{gulati2013comparative}. Here 50 pg/mL was significantly detected by our sensor and employed sensing method demonstrating its high sensitivity.

We extended our analysis by performing lock-in measurements at different frequencies, as shown in Fig.~\ref{fig_5}, to investigate the frequency-dependent response of the GFET biosensor. In Fig.~\ref{fig_5}a and Fig.~\ref{fig_5}b, we plot $X$, $Y$, and $Y/X^2$ as functions of $V_\mathrm {g}$ at 137~Hz and 1000~Hz, respectively, both before and after introducing the analyte. The flatter $X(V_\mathrm{g})$ curve at 1000~Hz compared to 137~Hz arises from the weakened electrolyte response, reduced effective gate control, and increased parasitic capacitance effects at higher frequencies. As a result, the charge carrier density in graphene is less modulated by $V_\mathrm{g}$, making the in-phase signal less dependent on gate voltage. This behavior aligns with the known frequency-dependent limitations of GFET's in electrolytes, as previously reported \cite{mackin2018frequency, garcia2020distortion}. At 137~Hz, ions in the electrolyte have sufficient time to redistribute in response to the gate voltage, allowing the formation of a strong electric double-layer effect and enabling efficient charge carrier modulation in graphene. In contrast, at 1000~Hz, the ionic movement cannot keep up with the fast oscillations, leading to a reduction in effective interfacial capacitance. This results in a weakened gate-induced doping effect, making the X vs. $V_{\mathrm g}$ curve appear less dependent on the applied gate voltage.

Further analyzing the effects of the analyte, we find that at 137~Hz, the leftward shift of the maximum in $X(V_{\mathrm g})$ confirms doping effects due to charge transfer between the analyte and graphene. However, at 1000~Hz, the overall $X$ decrease suggests that the analyte modifies the electrolyte-functional layer-electrode interface, leading to a reduction in effective gate capacitance rather than doping effects. This observation agrees with the well-documented frequency-dependent behavior of GFET's, where electrolyte dynamics dominate at low frequencies, while interfacial capacitance becomes significant at higher frequencies. In a typical EIS experiment for antibody-antigen biosensors, the system is analyzed over a range of frequencies to observe changes in $R_c$ and $C_{dl}$. At high frequencies, the impedance response is dominated by $C_{dl}$, which decreases upon antigen binding due to reduced ion mobility and changes in dielectric properties. At low frequencies, the impedance is controlled by $R_c$, which increases as the antigen layer blocks electron transfer \cite{lazanas2023electrochemical, randviir2022review}.

Previous EIS studies have used the Constant Phase Element (CPE) model to explain deviations from purely capacitive behavior of the electrolyte-graphene interface, which could potentially account for our observations as well. In \cite{sun2019unique}, the electrolyte-graphene interface was found to exhibit CPE behavior, with both the admittance parameter $Q_0$ and phase factor $\alpha$ varying with frequency. The study suggested that charged impurities and defects in the graphene lattice introduce inhomogeneous charge distribution, leading to frequency-dependent capacitance. Additionally, the low density of states of graphene near the Dirac point makes the phase factor particularly sensitive to local charge variations.

Similarly, in  \cite{garcia2020distortion}, a detailed characterization of liquid-gate GFET's frequency response revealed that signal distortion and transconductance reduction at higher frequencies arise due to deviations from the ideal-capacitor behavior. The study found that at low frequencies, ions in the electrolyte fully contribute to double-layer charging, enhancing the capacitance and gate modulation efficiency. However, at higher frequencies, ionic mobility constraints and capacitive leakage currents suppress the effective gate capacitance, causing weaker electrostatic control over the graphene channel.

With regard to our metric, by varying the measurement frequency while maintaining the same analyte concentration, we observed a substantial increase in the  $Y/X^2$ response at higher frequencies. This can be attributed to the fact that $Y/X^2$ reflects interfacial capacitance changes, which are more dominant at high frequencies, particularly upon antigen binding to the antibody. To further explore these effects, we conducted time-series measurements at different frequencies, tracking the evolution of $X$, $Y$, and $Y/X^2$ over time at a constant $V_\mathrm{g}$ corresponding to the peak transconductance while introducing the analyte (Fig.~\ref{fig_6}a). We observed larger shifts in $X$ and $Y/X^2$ for the same analyte concentration, reinforcing the dominant capacitive response at higher frequencies, as reflected in both $X$ and $Y/X^2$ measurements. In the high frequency regime, the response becomes capacitive-dominated, making the sensor highly sensitive to quantum capacitance and electric double layer ($C_\mathrm{dl}$) variations induced by analyte binding.

Interestingly, we also observed a reversal of the shift direction in $X$ at higher frequencies. This can be attributed to the non-ideal behavior of the electrolyte-graphene interface, where changes in interfacial capacitance shift the  $X(V_\mathrm{g})$ curve downward at higher frequencies (see Fig.~\ref{fig_5}). This behavior further highlights the role of CPE-like capacitance effects, where charge redistribution and interfacial capacitance become dominant, ultimately shaping the high-frequency response of the GFET biosensor.
 
Our GFET biosensor remains to be highly selective also at 1000~Hz, which was concluded from  comparison of sensor responses to the target- (PSA) and a non-target control (BSA) analytes (see Fig.~\ref{fig_6}b, and also Fig.~\ref{fig_4}a showing the results at 137~Hz). The sensor exhibited excellent selectivity, demonstrating a strong response to PSA, while the response to BSA was almost negligible. This confirms that the measured changes in both the resistive ($X$) and capacitive ($Y$) components, as well as the derived $Y/X^2$ ratio, are primarily due to the specific binding of the target analyte rather than non-specific interactions. The minimal response to BSA further supports that the observed capacitive changes at higher frequencies originate from PSA-antibody interactions, reinforcing the sensor's specificity and its effectiveness in distinguishing bio-molecular interactions through $Y/X^2$.

Finally, we compared the $Y/X^2$ response at 1000~Hz to a conventional DC measurement of resistance (see Fig.~\ref{fig_6}c). Both DC and AC measurements were performed on the same device using the same analyte, ensuring a direct comparison between the two approaches. The results clearly demonstrate that the lock-in measurement response far exceeds what can be detected using conventional DC methods. The significantly higher sensitivity observed in the lock-in measurement highlights the advantage of frequency-based detection, where enhanced capacitive contributions and improved noise rejection enable the detection of subtle bio-molecular interactions that remain undetectable in conventional resistance-based sensing. This confirms that $Y/X^2$ at higher frequencies provides a superior biosensing strategy, offering enhanced sensitivity and robustness over traditional methods.

\section{\label{sec:level1} Conclusion}

We have introduced a complementary electrochemical impedance spectroscopy (EIS) method for biosensing. In this innovative approach, the AC bias current is applied through a resistive thin film that is in contact with both the bio-functional layer and the electrolyte, rather than through the electrolyte itself as in traditional EIS. The in-phase ($X$) and quadrature ($Y$) components of the AC voltage across the thin film provide valuable information about the bio-functional layer.

The bio-functional layer can be engineered to undergo changes upon molecular bindings, similar to classical antibody-antigen interactions. A new and useful metric of this method, denoted as 
$Y/X^2$, has been identified. This metric effectively isolates changes in the capacitance of the bio-functional layer and the adjacent double layer of the electrolyte.

We have demonstrated the effectiveness of this method using graphene-liquid-gate-transistor biosensors. In our experiments, specific antibody-antigen binding was reliably detected even in analytes with very low concentrations of antigen. The sensor response was notably free from noise and the baseline drift that is commonly observed in graphene biosensors utilizing conventional DC resistance measurements.

Most importantly, this method is not limited to graphene biosensors. It can be broadly applied to any conducting thin films with sufficiently high sheet resistance. This versatility expands the range of potential biosensing platforms by allowing for the customization of chemical linkers between the resistive-film material and antibodies or other functional molecules.
 
\begin{acknowledgments}
{This research has received funding from the European Union’s Horizon 2020 research and innovation programme under the Marie Skłodowska-Curie grant agreement No 955626 and Nordic Programme for Interdisciplinary Research, grant 105121. Support from 2D-TECH (Vinnova) competence center is highly appreciated. SP acknowledges funding from Vetenskapsrådet (2020–04096). This work was performed in part at Myfab Chalmers.}
\end{acknowledgments}

\bibliography{aapmsamp}

\end{document}